\documentclass[sigconf,screen, natbib=true]{acmart}
\usepackage{graphicx}
\usepackage{subcaption}
\usepackage{algorithm}
\usepackage{algpseudocode}
\usepackage{enumitem}
\usepackage{multirow}
\usepackage{tabularx}
\usepackage{threeparttable}
\usepackage{color, xspace}
\usepackage{appendix}
\usepackage{balance}

\AtBeginDocument{%
	\providecommand\BibTeX{{%
			\normalfont B\kern-0.5em{\scshape i\kern-0.25em b}\kern-0.8em\TeX}}}
\copyrightyear{2025}
\acmYear{2025}
\setcopyright{acmlicensed}\acmConference[KDD '25]{Proceedings of the 31st ACM
SIGKDD Conference on Knowledge Discovery and Data Mining V.2}{August 3--7,
2025}{Toronto, ON, Canada}
\acmBooktitle{Proceedings of the 31st ACM SIGKDD Conference on Knowledge
Discovery and Data Mining V.2 (KDD '25), August 3--7, 2025, Toronto, ON, Canada}
\acmDOI{10.1145/3711896.3736968}
\acmISBN{979-8-4007-1454-2/2025/08}

\begin{document}
\title[FindRec: Stein-Guided Entropic Flow for Multi-Modal Sequential Recommendation]{FindRec: Stein-Guided Entropic Flow for  \\ Multi-Modal Sequential Recommendation}

\author{Maolin Wang}
\authornote{Equal contribution.}
\affiliation{%
  \institution{City University of Hong Kong}
  \streetaddress{}
  \city{Hong Kong SAR}
  \state{}
  \country{China}
}
\email{morin.wang@my.cityu.edu.hk}

\author{Yutian Xiao}
\affiliation{%
  \institution{Beihang University}
  \streetaddress{}
  \city{Beijing}
  \state{}
  \country{China}
}
\email{by2442221@buaa.edu.cn}

\author{Binhao Wang}
\authornotemark[1] 
\affiliation{%
  \institution{City University of Hong Kong}
  \streetaddress{}
  \city{Hong Kong SAR}
  \state{}
  \country{China}
}
\email{binhawang2-c@my.cityu.edu.hk}

\author{Sheng Zhang}
\affiliation{%
  \institution{City University of Hong Kong}
  \streetaddress{}
  \city{Hong Kong SAR}
  \state{}
  \country{China}
}

\email{szhang844-c@my.cityu.edu.hk}

\author{Shanshan Ye}
\affiliation{%
  \institution{University of Technology Sydney}
  \streetaddress{}
  \city{Sydney}
  \state{New South Wales}
  \country{Australia}
  \postcode{}
}
\email{shanshan.ye@student.uts.edu.au}

\author{Wanyu Wang}
\authornote{Corresponding author.}
\affiliation{%
  \institution{City University of Hong Kong}
  \streetaddress{}
  \city{Hong Kong SAR}
  \state{}
  \country{China}
  \postcode{}
}
\email{wanyu.wang@my.cityu.edu.hk}

\author{Hongzhi Yin}
\affiliation{%
  \institution{The University of Queensland}
  \streetaddress{}
  \city{Brisbane}
  \state{Queensland}
  \country{Australia}
  \postcode{}
}
\email{db.hongzhi@gmail.com}

\author{Ruocheng Guo}
\affiliation{%
  \institution{Independent Researcher}
  \streetaddress{}
  \city{Hong Kong SAR}
  \state{}
  \country{China}
  \postcode{}
}
\email{rguo.asu@gmail.com}

\author{Zenglin Xu}
\affiliation{%
  \institution{Fudan University}
  \streetaddress{}
  \city{Shanghai}
  \state{}
  \country{China}
  \postcode{}
}
\email{zenglin@gmail.com}



\renewcommand{\shortauthors}{Maolin Wang, et al.}
\begin{abstract}
Modern recommendation systems face significant challenges in processing multimodal sequential data, particularly in temporal dynamics modeling and information flow coordination. Traditional approaches struggle with distribution discrepancies between heterogeneous features and noise interference in multimodal signals. We propose \textbf{FindRec}~ (\textbf{F}lexible unified \textbf{in}formation \textbf{d}isentanglement for multi-modal sequential \textbf{Rec}ommendation), introducing a novel "information flow-control-output" paradigm. The framework features two key innovations: (1) A Stein kernel-based Integrated Information Coordination Module (IICM) that theoretically guarantees distribution consistency between multimodal features and ID streams, and (2) A cross-modal expert routing mechanism that adaptively filters and combines multimodal features based on their contextual relevance. Our approach leverages multi-head subspace decomposition for routing stability and RBF-Stein gradient for unbiased distribution alignment, enhanced by linear-complexity Mamba layers for efficient temporal modeling. Extensive experiments on three real-world datasets demonstrate FindRec's superior performance over state-of-the-art baselines, particularly in handling long sequences and noisy multimodal inputs. Our framework achieves both improved recommendation accuracy and enhanced model interpretability through its modular design. The implementation code is available anonymously online for easy reproducibility~\footnote{https://github.com/Applied-Machine-Learning-Lab/FindRec}.

\end{abstract}




\maketitle
\section{Introduction}
\label{sec:introduction}
Modern recommendation systems face unprecedented challenges in processing complex multimodal user behavior data, where user interactions naturally encompass both multimodal signals (e.g., text and images) and temporal dynamics (e.g., long-term preferences and short-term interests)~\cite{jing2023contrastive,hu2023adaptive}. While traditional ID-based sequential models effectively capture basic interaction patterns~\cite{wang2019sequential}, they struggle with two critical challenges that limit their real-world effectiveness. First, temporal dynamics modeling remains insufficient, as user interests rapidly evolve with time and context~\cite{ye2020time}, requiring simultaneous capture of both long-term trends and short-term fluctuations. Second, information flow coordination faces fundamental challenges due to the dual complications of distribution discrepancy and noise interference. Specifically, the significant distribution gap between heterogeneous multimodal features (visual-textual signals) and sequential behaviors (ID interaction streams) introduces systematic integration bias, while substantial irrelevant information in product descriptions (e.g., packaging images and usage instructions) further masks genuine interest signals and impedes model interpretability and performance~\cite{wang2024aligned,liang2023mmmlp}.

These challenges manifest prominently in sequential recommendation scenarios, where multimodal signals are inherently complex and noisy~\cite{vasile2016meta}. Product images often contain decorative elements or promotional materials that do not reflect user preferences and textual descriptions can mix essential attribute information with generic marketing content~\cite{li2023text}. Moreover, user interests demonstrate complex temporal patterns, including both stable components (such as brand loyalty and category preferences) and dynamic elements (like seasonal trends and contextual needs), which exceed the capabilities of conventional temporal modeling approaches~\cite{fan2021continuous}.

Investigating the complexities of sequential user behavior and evolving preferences, the field has explored a range of modeling approaches. Deep reinforcement learning, for instance, has emerged as a prominent direction, framing recommendation as a sequential decision-making process~\cite{zhao2019deep, zhao2018deep, zhao2018recommendations, zhao2021dear, zhao2023user}. Concurrently, significant efforts in multimodal recommendation have sought to tackle the aforementioned challenges through various approaches~\cite{jing2023contrastive}.Early fusion methods directly concatenate multimodal features (e.g., MV-RNN)~\cite{cui2018mv}, but suffer from noise propagation that corrupts the overall representation. Adaptive fusion approaches, such as attention-based (wang2023missrec)~\cite{wang2023missrec}, mixture-of-experts based (M3oE) 
~\cite{zhang2024m3oe}, or hierarchical time-aware experts based~(HM4SR) mechanisms~\cite{zhang2025hierarchical}, dynamically weight different modalities but remain sensitive to distribution shifts. Temporal enhancement methods leverage graph neural networks or frequency-domain transformations (FEARec)~\cite{du2023frequency} to model sequential patterns, yet rely on manually designed fusion stages. These existing solutions exhibit three critical limitations: (1) lack of alignment between multimodal and ID streams leading to information conflicts~\cite{wang2024aligned}, and (2) varying contribution importance across modalities, where aligned features have different degrees of relevance that static fusion methods cannot adaptively handle.
 
\begin{figure*}[t]
    \centering
    \includegraphics[width=\linewidth]{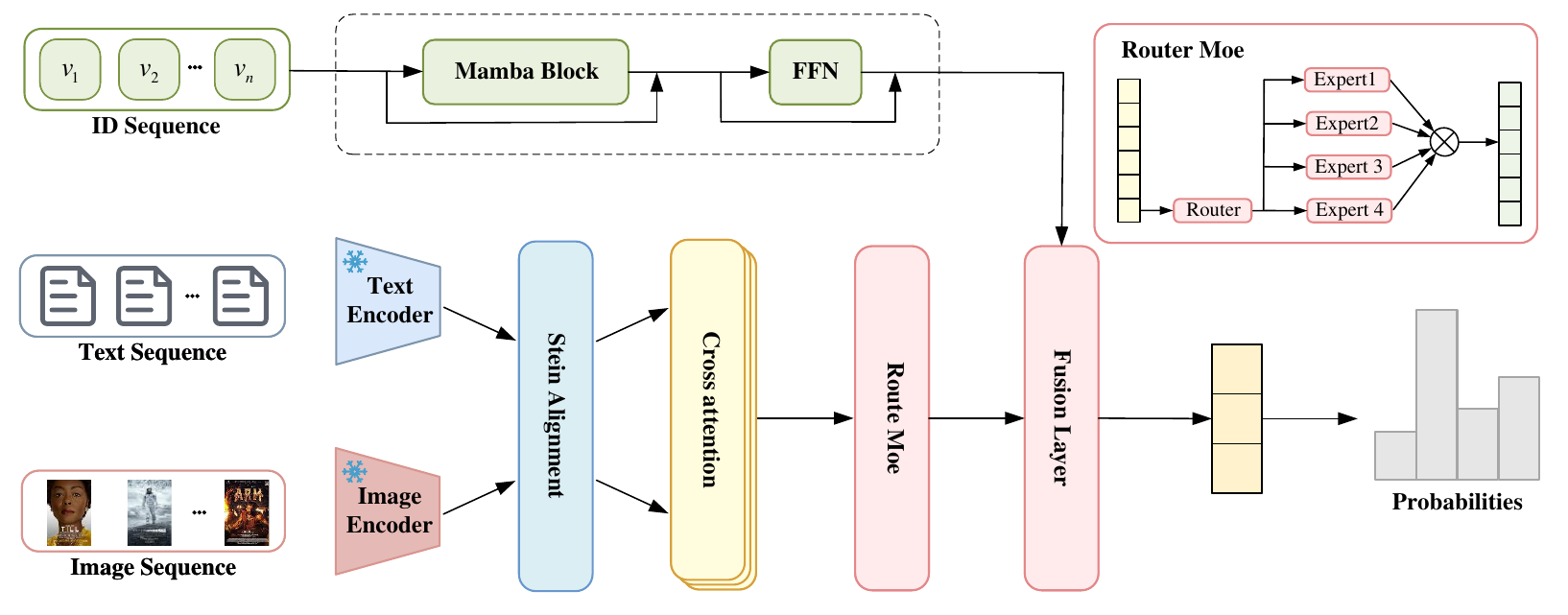}
    \caption{The overall architecture of FindRec. The framework processes three types of input sequences (ID sequence, text sequence, and image sequence) through multiple processing stages including Mamba-FFN, modality-specific encoders, Stein alignment, and router-based expert mechanism to generate final prediction probabilities. The bottom panels illustrate the detailed designs of Stein alignment mechanism (left) and router MoE architecture (right).}
    \label{fig:main}
\end{figure*}
To address these fundamental limitations, we propose \textbf{F}lexible {U}nified \textbf{in}formation \textbf{d}isentanglement for {M}ulti-{M}odal sequential \textbf{Rec}ommendation (\textbf{FindRec}), introducing a novel \textit{information flow-control-output} paradigm.
FindRec addresses technical challenges through multi-head subspace decomposition for routing stability and RBF-Stein gradient~\cite{liu2016stein} (a kernel-based method~\cite{wen2021gradient,wen2020mutual,liu2016stein,liu2021leveraging,wen2024mveb} that combines radial basis functions with Stein's operator for accurate gradient approximation) for unbiased distribution alignment. To enhance the temporal dependencies modeling, we leverage linear-complexity Mamba layers~\cite{gu2023mamba,liu2024mamba4rec,qu2024survey,jiang2025dual} which provide efficient sequential processing through state space models while maintaining theoretical guarantees for distribution alignment.
Following this paradigm, our framework achieves superior recommendation performance while maintaining strong model interpretability. FindRec has two key novelties: (1) A Stein kernel-based Integrated Information Coordination Module (IICM) that theoretically guarantees distribution consistency between multimodal features and ID streams; and (2) A cross-modal expert routing mechanism that adaptively filters and combines multimodal features based on their varying degrees of relevance, addressing the challenge of unequal contribution importance.

The main contributions are summarized as follows:
\begin{itemize}[leftmargin=*]
\item \textbf{Stein-Enhanced Multimodal Alignment}: To address the \textit{distribution inconsistency challenge}, our IICM module leverages Stein kernel-based synchronization and differential entropy maximization, achieving provable cross-modal consistency while preserving modality-specific information. This theoretically guarantees unbiased distribution alignment between multimodal features and ID feature streams.

\item \textbf{Dynamic Information Flow Control}: To tackle the \textit{relevance assessment challenge}, we propose an adaptive expert routing mechanism that dynamically filters and combines multimodal features based on their contextual importance. This is further enhanced by a novel multi-head subspace decomposition approach for routing stability, effectively handling varying degrees of feature contribution in cross-modal fusion.

\item \textbf{Extensive Empirical Validation}: Through comprehensive experiments on three real-world datasets (MovieLens-100k, Micro-lens, Amazon Beauty), we demonstrate FindRec's consistent performance gains over state-of-the-art baselines, including recent multimodal sequential recommenders.

\end{itemize}

\section{Methodology}

In this section, we present FindRec, a novel recommendation framework that systematically integrates multimodal information with temporal modeling. Following an ``information flow-control-output'' paradigm, our framework achieves superior recommendation performance while maintaining strong model interpretability. As shown in Figure~\ref{fig:main}, the system architecture of {FindRec} follows a clear data flow path: first extracting key information from item IDs, text, and images and projecting them into a unified latent space; then capturing users' long-term preferences and short-term dynamics through hierarchical temporal modeling; followed by dynamic selection and integration of multimodal signals via a cross-modal expert routing module; further alignment and control of auxiliary signals through an integrated information coordination module; and finally fusing all components for final prediction. This design directly addresses quality assurance of information flow and interpretability control while enabling robust handling of multi-modal recommendations.
\vspace{-1mm}
\subsection{Feature Extraction and ID Modeling}
Traditional recommendation systems face several challenges~\cite{zhou2023comprehensive}: (1) relying solely on item IDs fails to capture rich multimodal information that influences user preferences, (2) existing multimodal fusion approaches often treat different modalities independently, leading to suboptimal feature interactions, and (3) modeling complex temporal dependencies in sequential user behaviors remains difficult, especially when integrating multimodal signals.

To address these challenges, we design a comprehensive multimodal feature extraction module. For each item, we extract and project modality-specific features using pre-trained models: item embeddings $e_{id} = \mathrm{ItemEmbedding}(v)$ capture inherent item characteristics, text embeddings $e_{txt} = W_{txt}\cdot f_{text}(t)$ encode semantic information, and vision features $e_{img} = W_{img}\cdot f_{ViT}(I)$ extract visual patterns, all projected to dimension $d$. These features are fused through concatenation to form $e_{fused} = \mathrm{Concat}(e_{id}, e_{txt}, e_{img})$.

To effectively model temporal dynamics, {FindRec} employs Mamba-based state-space models~\cite{gu2023mamba}, which overcome the limitations of traditional transformers in capturing both fine-grained patterns and long-range dependencies. While transformers rely on self-attention that scales quadratically with sequence length and may struggle to capture precise temporal patterns, Mamba's state space modeling provides linear complexity and better handles continuous-time dynamics. Given the item embedding sequence $L^{(embed)}$, our model obtains temporal representations through $z_{ID} = \mathrm{MambaLayer}(L^{(embed)})$. Each Mamba layer follows the transformation:
$x_{\ell+1} = x_\ell + \alpha\cdot (\mathrm{Mamba}(\mathrm{LayerNorm}(x_\ell)))$
where $\mathrm{Mamba}(\cdot)$ implements the state space model by:
$\hat{x} = \Delta \odot f_{\Delta}(x) + f_B(x)$
$h_{t} = \mathrm{SSM}(\hat{x})$
$y = h_{t} \odot f_D(x)$
This transformation incorporates LayerNorm for stability, Mamba operations for temporal dependency capture, and dropout for regularization. The learnable scaling parameter $\alpha$ controls information flow between layers. The multi-scale temporal patterns are captured through the hierarchical processing of the SSM and residual connections: while the SSM component models local dependencies through state transitions, the residual connections preserve and accumulate information across different temporal scales, allowing the model to capture both short-term dynamics and long-term trends.

\begin{figure}[t]
\centering
\setlength{\fboxrule}{0.pt}
\setlength{\fboxsep}{0.pt}
\fbox{ \includegraphics[width=0.9\linewidth]{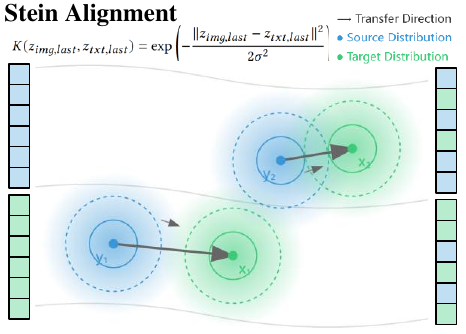}
}
\caption{Visualization of Stein alignment mechanism in IICM. The diagram shows how multimodal feature vectors are aligned through RBF kernel function $k(x,y)=\exp(-|x-y|^2/h)$, where concentric circles represent the kernel's influence regions. Blue dots ($y_1$, $y_2$) indicate sample points in the feature space, while the red one represents the target point for alignment. The rectangles on both sides illustrate the feature embeddings from different modalities being aligned.}
\label{stein}
\vspace{-6mm}
\end{figure}

\subsection{Stein-based Integrated Information Coordination}
The multimodal signals must be precisely calibrated and aligned before merging with the primary ID sequence. Our Integrated Information Coordination Module (IICM) is designed to govern and regulate the flow of information using a combination of Stein kernel~\cite{liu2016stein,wen2024mveb} based similarity estimation and KL divergence regularization, ensuring that the embedded representations from two modalities branches remain both consistent and informative throughout the sequential recommendation pipeline.

Our work leverages Stein methods's unique ability to capture complex dependencies and perform flexible distribution matching. Compared to traditional contrastive learning approaches that rely on sample-based negative pairs, Stein kernel methods offer several key advantages: (1) They directly optimize distribution alignment without requiring careful negative sampling strategies or large batch sizes, making training more stable and efficient; (2) While contrastive learning may suffer from representation collapse or modal dominance issues, Stein methods naturally preserve the geometric structure of each modality's feature space through their kernel formulation; and (3) The score-based gradient estimation provides an unbiased way to maximize feature entropy, preventing the ``shortcut learning'' problem where models may exploit simple statistical correlations (such as dominant colors in images or high-frequency words in text) rather than learning semantically meaningful cross-modal relationships. This is particularly crucial for recommendation scenarios where we need to capture subtle but important multimodal patterns beyond superficial similarities.

Initially, as demonstrated in Figure~\ref{stein}, the module extracts the final nonlinear representations $z_{img,last}$ and $z_{txt,last}$ from the respective branches after extensive processing. These vectors encapsulate the distilled semantic information of each modality. To quantify the similarity between these modal embeddings, we employ a Radial Basis Function (RBF) kernel~\cite{liu2016stein}, which is expressed as
\begin{equation}
K(z_{img,last}, z_{txt,last}) = \exp\left(-\frac{\|z_{img,last} - z_{txt,last}\|^2}{2\sigma^2}\right),
\end{equation}
where $\sigma$ is an adaptively estimated bandwidth parameter that is updated via the Stein kernel mechanism. This non-linear similarity measure captures complex dependencies between the modalities that simpler linear metrics would miss. The average similarity across the batch is then defined as the alignment loss
\begin{equation}
\mathcal{L}_{IICM} = \mathbb{E}\Bigl[K(z_{img,last}, z_{txt,last})\Bigr],
\end{equation}
which encourages the representations from both modalities to converge toward a coherent, unified space while preserving their unique discriminative features.

A key design highlight of our IICM is its dynamic control over the information flow. The Stein kernel not only facilitates the computation of $K(z_{img,last}, z_{txt,last})$ but also provides an unbiased, score-based estimation of the entropy gradients with respect to the model parameters. In essence, the Stein gradient estimator~\cite{liu2016stein,wen2024mveb} taps into the local geometry of the feature distributions by approximating the score function $\nabla_z \log q(z)$. This approximation is crucial for maximally increasing the differential entropy of the embeddings, thus promoting global uniformity across latent space.

By enforcing this dual objective—maximizing alignment while simultaneously regularizing the latent distributions—the IICM serves as a robust ``gatekeeper'' in our framework. The alignment process operates through a two-stage mechanism: first, the Stein kernel-based synchronization actively minimizes the distributional discrepancy between modality pairs, while the differential entropy regularization term prevents over-alignment and preserves modality-specific characteristics. This careful balance ensures that only high-quality, consistently aligned information flows through to subsequent stages, effectively filtering out noise and irrelevant signals.

\subsection{Cross-modal Attention and Expert Routing}
While Stein alignment provides calibrated features, effectively utilizing features with varying importance remains challenging. Different aspects of aligned features contribute unequally to the final representation - some features carry more relevant information, while others are less important for the current context. Traditional static fusion methods cannot adaptively handle such varying degrees of contribution importance. To address these challenges, we propose a cross-modal expert routing module that dynamically filters features based on their relevance from $z_{img,last}$ and $z_{txt,last}$.

The module divides these aligned representations into $H$ subspaces (attention heads), where each head processes a low-dimensional representation $d_h=d/H$. For each head, we compute cross-attention scores $A^h = \mathrm{Softmax}(z^h_{img,last}(z^h_{txt,last})^\top/\sqrt{d_h})$ between image and text subvectors, then combine them through $O^h = z^h_{img,last} + A^h\cdot z^h_{txt,last}$ to capture cross-modal interactions. The cross-attention mechanism enables each modality to attend to relevant aspects of the other modality - when computing $A^h$, the dot product between $z^h_{img,last}$ and $(z^h_{txt,last})^\top$ measures the compatibility between image and text features, essentially learning which parts of the text are most relevant to each image region and vice versa. The scaling factor $\sqrt{d_h}$ ensures numerical stability during training by preventing the dot products from growing too large in magnitude, which could lead to extremely peaked softmax distributions~\cite{liu2016large,vaswani2017attention}.

We employ a lightweight expert router that computes gating weights $g^h$ for each head's output to achieve dynamic information control. While standard attention mechanisms excel at capturing general dependencies, they may struggle with the diverse and specialized patterns in cross-modal data. The expert routing mechanism addresses this limitation by introducing specialized processing pathways. The router determines the optimal combination through expert aggregation $E^h = \sum_{k=1}^{N_e} g^h_k\cdot \mathrm{Expert}k(O^h)$, followed by concatenation $O{exp} = \mathrm{Concat}(E^1,\dots,E^H)$ across heads. The expert mechanism is particularly powerful as each $\mathrm{Expert}_k$ specializes in processing different types of cross-modal patterns - some experts might focus on high-level semantic relationships while others capture fine-grained details or modality-specific nuances. The gating weights $g^h_k$ act as learned importance scores, adaptively and dynamically routing information through the most appropriate experts based on the contextual input features and modal characteristics.

This adaptive routing effectively manages information flow by identifying valuable cross-modal signals while mitigating noise interference.
The processed cross-modal representations $O_{exp}$ serve as enhanced complementary signals for subsequent recommendation tasks. By combining the strengths of cross-attention and expert routing, our module ensures that the final representations maintain the high-quality alignments established by the Stein kernel~\cite{bishop2006pattern,liu2016stein,wen2024mveb} while further refining the cross-modal interactions through specialized expert processing.

\subsection{Fusion and Prediction}
Effectively integrating temporal sequential patterns with multimodal information presents several key challenges: (1) naive fusion methods often lead to information interference and degraded performance, (2) maintaining interpretability while combining complex temporal and multimodal signals is non-trivial, and (3) balancing the contribution of each information stream requires careful calibration to prevent one modality from dominating the others.
{FindRec} employs a carefully designed integration strategy in the final stage to address these challenges. We combine the temporal behavior patterns $z_{ID}$ from Mamba layers with the refined multimodal features $O_{exp}$ from expert routing and coordination modules through a structured concatenation approach:
\begin{align}
z_{final} &= \mathrm{Concat}(z_{ID}, O_{exp}), \\
S &= \mathrm{FFN}(z_{final}),
\end{align}
The concatenated representation undergoes deep feature extraction through a feed-forward network (FFN) to produce the final sequence embedding. This architecture allows the model to learn complex interactions between temporal patterns and multimodal features while maintaining the interpretability of each component's contribution.
During training, we jointly optimize the recommendation loss $\mathcal{L}_{rec}$ (e.g., BPR or cross-entropy loss) and the IICM loss, with the overall objective defined as:
\begin{align}
L(\phi) = \mathcal{L}_{rec} + \lambda\,\mathcal{L}_{IICM}
\end{align}
 The recommendation loss focuses on the primary task of accurate item prediction, while the IICM loss ensures proper alignment and regularization of multimodal features. This joint optimization strategy enables the model to learn high-quality representations that balance task performance with representation quality.
\section{Experiments}

To validate the performance of our FindRec framework, we designed and conducted comprehensive experimental studies. In this section, we present our experimental setup and results analysis. Specifically, our experiments aim to investigate the following research questions:

\begin{itemize}[leftmargin=*]
\item \textbf{RQ1:} How does FindRec perform compared to state-of-the-art recommendation models across different dataset scales?
\item \textbf{RQ2:} How does our multimodal information coordination mechanism perform with different sequence lengths, especially for short sequences with limited behavioral signals?
\item \textbf{RQ3:} What are the individual contributions of the cross-modal expert routing and integrated information coordination modules?
\item \textbf{RQ4:} How do key hyperparameters in the information flow control mechanism affect model performance?
\end{itemize}

\subsection{Datasets and Evaluation Protocol}

We conduct comprehensive experiments on three diverse datasets: MovieLens-100k, Micro-lens, and Amazon Beauty, each incorporating both sequential interaction data and rich multimodal information.
MovieLens-100k contains about 100k ratings from 944 users on 1,334 movies, where each interaction includes user ID, movie ID, rating (1-5), and timestamp. For multimodal enhancement, we collect movie posters and textual metadata including titles, genres, and plot summaries. The dataset features 98,609 interactions with an average sequence length of 104.57 and sparsity of 92.17\%.

Micro-lens~\cite{ni2025content} is a concise dataset specifically processed for multimodal sequential recommendation, containing 98,130 users, 17,229 items, and 705,174 interactions. Each micro-video is associated with high-resolution poster images (224×224 pixels) and rich textual descriptions averaging 128 tokens in length. The interaction density is 99.96\% with an average of 7.19 interactions per user.

Amazon Beauty provides a diverse e-commerce scenario, enhanced with product images and textual descriptions. It encompasses 60,276 interactions with 4,323 users and 2,424 items. Each product includes standardized images (224×224 resolution) and detailed textual descriptions averaging 256 words, with a sparsity rate of 99.42\%. The dataset statistics are summarized in Table~\ref{sec:statistics}.


In data preprocessing, we follow standard practices in sequential recommendation. Users with fewer than 5 interactions are filtered out to ensure sufficient sequential patterns. Each user's interaction sequence is chronologically sorted and split with a ratio of 8:1:1, allocating 80\% of interactions to training, 10\% to validation, and 10\% to testing. This ensures a realistic temporal evaluation while maintaining consistent splits across all datasets. All models, including the reproduced SASRec baseline, are trained and evaluated under this identical protocol using RecBole. For multimodal feature extraction, images are encoded by BLIP~\cite{li2022blip} and text is encoded by RoBERTa~\cite{liu2020roberta}.

\begin{table}[t]
\centering
\caption{Statistics of the evaluation datasets}
\label{sec:statistics}
\resizebox{0.48\textwidth}{!}{
\begin{tabular}{lccccc}
\hline
Dataset & Users & Items & Interactions & Sparsity & Avg. Seq. Length \\
\hline
Micro-lens & 98130 & 17229 & 705174 & 99.96\% &  7.19\\
MovieLens-100k & 944 & 1334 & 98609 & 92.17\% & 104.57 \\
Amazon Beauty & 4323 & 2424 & 60276 & 99.42\% & 13.95 \\
\hline
\end{tabular}
}
\end{table}

\begin{table*}[t]
    \caption{Performance comparison on three real-world datasets. Methods are categorized into (1) ID-based sequential methods that only utilize interaction sequences, (2) Simple multi-modal methods that directly incorporate multi-modal features, and (3) Advanced multimodal methods with sophisticated fusion mechanisms. \textbf{Bold} numbers denote the best performance, \underline{underlined} numbers represent the second-best results, and $^\ast$ indicates statistical significance at p < 0.05 level using a paired t-test. The bottom row shows the relative improvements of our method over the best baseline.}
    \label{tab:performance}
    \renewcommand{\arraystretch}{1.05}
    \resizebox{\linewidth}{!}{
    \begin{tabular}{ccccccccccccc}
        \toprule
        \multirow{2}{*}{Methods} 
        & \multicolumn{4}{c}{MicroLens} 
        & \multicolumn{4}{c}{MovieLens-100K} 
        & \multicolumn{4}{c}{Amazon Beauty} \\
        \cline{2-13}
        & NDCG@5 & NDCG@10 & MRR@5 & MRR@10
        & NDCG@5 & NDCG@10 & MRR@5 & MRR@10
        & NDCG@5 & NDCG@10 & MRR@5 & MRR@10\\
        \midrule
        \multicolumn{13}{l}{\textbf{ID-based Sequential Methods:}} \\
        SASRec 
        & 0.0352 & 0.0428 & 0.0287 & 0.0323
        & 0.1835 & 0.2311 & 0.1478 & 0.1646
        & 0.0635 & 0.0796 & 0.0525 & 0.0584\\
        BERTRec 
        & 0.0361 & 0.0439 & 0.0294 & 0.0331
        & 0.1847 & 0.2365 & 0.1512 & 0.1639
        & 0.0643 & 0.0804 & 0.0536 & 0.0598\\
        GRU4Rec 
        & 0.0355 & 0.0434 & 0.0291 & 0.0325
        & 0.1789 & 0.2292 & 0.1482 & 0.1629
        & 0.0627 & 0.0781 & 0.0521 & 0.0578\\
        SMLP4Rec 
        & 0.0378 & 0.0458 & 0.0309 & 0.0347
        & 0.1871 & 0.2436 & 0.1534 & 0.1709
        & 0.0653 & 0.0816 & 0.0543 & 0.0606\\
        Mamba4Rec 
        & 0.0415 & 0.0507 & 0.0335 & 0.0381
        & 0.2193 & 0.2723 & 0.1756 & 0.1987
        & 0.0781 & 0.0948 & 0.0671 & 0.0735\\
        \midrule
        \multicolumn{13}{l}{\textbf{Simple Multi-modal Sequential Methods:}} \\
        SASRec~(MM) 
        & 0.0342 & 0.0415 & 0.0278 & 0.0312
        & 0.1780 & 0.2242 & 0.1434 & 0.1597
        & 0.0616 & 0.0772 & 0.0509 & 0.0567\\
        Mamba4Rec~(MM) 
        & 0.0402 & 0.0492 & 0.0325 & 0.0369
        & 0.2127 & 0.2641 & 0.1703 & 0.1927
        & 0.0758 & 0.0920 & 0.0651 & 0.0713\\
        \midrule
        \multicolumn{13}{l}{\textbf{Advanced Multi-modal Sequential Methods:}} \\
        NOVA 
        & 0.0423 & 0.0522 & 0.0358 & 0.0397
        & 0.2288 & 0.2792 & 0.1864 & 0.2097
        & 0.0817 & 0.0995 & 0.0696 & 0.0771\\
        UniSRec 
        & 0.0420 & 0.0519 & 0.0353 & 0.0393
        & 0.2281 & 0.2795 & 0.1862 & 0.2100
        & 0.0815 & 0.0986 & 0.0683 & 0.0759\\
        IISAN 
        & 0.0421 & 0.0524 & 0.0357 & 0.0399
        & 0.2285 & 0.2791 & 0.1871 & 0.2104
        & 0.0819 & 0.0998 & 0.0698 & 0.0773\\
        MMMLP 
        & \underline{0.0425} & \underline{0.0527} & \underline{0.0359} & \underline{0.0401}
        & \underline{0.2293} & \underline{0.2801} & \underline{0.1877} & \underline{0.2112}
        & \underline{0.0822} & \underline{0.1002} & \underline{0.0700} & \underline{0.0776}\\
        \midrule
        \textbf{Ours} 
        & \textbf{0.0438}$^\ast$ & \textbf{0.0540}$^\ast$ & \textbf{0.0371}$^\ast$ & \textbf{0.0408}$^\ast$ 
        & \textbf{0.2325}$^\ast$ & \textbf{0.2830}$^\ast$ & \textbf{0.1933}$^\ast$ & \textbf{0.2165}$^\ast$
        & \textbf{0.0843}$^\ast$ & \textbf{0.1027}$^\ast$ & \textbf{0.0722}$^\ast$ & \textbf{0.0797}$^\ast$\\
        \midrule
        Improv. 
        & 3.06\% & 2.47\% & 3.34\% & 1.75\% 
        & 1.40\% & 1.04\% & 2.98\% & 2.51\% 
        & 2.55\% & 2.50\% & 3.14\% & 2.71\%\\
        \bottomrule
    \end{tabular}}
\end{table*}
\subsection{Baseline Methods}
We compare our model with several representative baseline methods, which can be categorized into three groups:

\noindent\textbf{ID-based Sequential Recommendation Methods}: 
1) SASRec \cite{kang2018self} captures long-term and short-term user preferences by applying a multi-head attention mechanism to model user behavior sequences adaptively. 
2) BERTRec \cite{sun2019bert4rec} adapts the Bidirectional Encoder Representations from Transformers (BERT) architecture to model user behaviors for personalized recommendation. 
3) GRURec \cite{hidasi2015session} utilizes GRUs to capture sequential dependencies within user interactions for session-based recommendations. 
4) SMLP4Rec \cite{gao2024smlp4rec} employs a tri-directional fusion scheme to learn correlations on sequence, channel, and feature dimensions efficiently. 
5) Mamba4Rec \cite{liu2024mamba4rec} explores the potential of selective SSMs for efficient sequential recommendation, substantially improving SRS models' efficiency.

\noindent\textbf{Simple Multi-modal Sequential Methods}: 
1) SASRec~(MM) is the basic multi-modal SASRec that directly mixes modal features into ID embeddings, and we simply feed multi-modal signals into the original SASRec network as additional features.
2) Mamba4Rec~(MM) integrates multi-modal features into the Mamba architecture to explore the application of SSM in multi-modal sequential recommendation, and similarly, we incorporate multi-modal signals directly into the Mamba backbone network.

\noindent\textbf{Advanced Multi-modal Sequential Methods}: 
1) NOVA \cite{liu2021noninvasive} leverages product images and textual descriptions through vision-text contrastive learning to enhance sequence representation for better recommendation performance. 
2) UniSRec \cite{hou2022towards} presents a unified multi-modal sequential recommender framework that seamlessly integrates images, text, and user behavior data, learning their relationships in a unified manner. 
3) IISAN \cite{fu2024iisan} is a multi-modal recommendation method based on item-item self-attention networks, which specifically focuses on capturing multi-modal similarity relationships between items. 
4) MMMLP \cite{liang2023mmmlp}, a purely MLP-based architecture, processes multi-modal data through three key modules: Feature Mixer Layer, Fusion Mixer Layer, and Prediction Layer, achieving state-of-the-art performance with linear complexity.

\subsection{Implementation Details}
In this subsection, we introduce the implementation details of FindRec.\footnote{Some hyperparameter values in the camera-ready version contain typos; the values reported here follow the corrected arXiv version.} We employ the AdamW optimizer~\cite{loshchilov2017decoupled} with a learning rate of 0.002 for model training. The training and evaluation processes utilize a batch size of 256. The item ID embedding dimension is set to 128 for MicroLens and 64 for both Amazon Beauty and ML-100k. For multimodal features, image features are extracted by BLIP~\cite{li2022blip} and text features are extracted by RoBERTa~\cite{liu2020roberta}; both are projected to a hidden dimension of 512 to ensure balanced representation learning. Considering the varying sequence characteristics shown in Table~\ref{sec:statistics}—where MicroLens, Beauty, and ML-100k have average sequence lengths of 7.19, 13.95, and 104.57, respectively—we configure the maximum sequence length as 50 for MicroLens and Beauty, and extend it to 100 for ML-100k. To address the data sparsity in MicroLens and Amazon Beauty, we implement a dropout rate of 0.3, compared to 0.2 for ML-100k. For mamba-based components, we incorporate two mamba layers to optimize their performance. The model is trained for 100 epochs with cosine learning rate scheduling and 2000 warmup steps, and gradient clipping with a threshold of 1.0 is applied for training stability. For mamba-based baseline models, the remaining implementation details align with the configurations from the original papers. We adopt NDCG and MRR at top-K values of 5 and 10 as evaluation metrics. All experiments were conducted on an NVIDIA RTX 4090 GPU with 10 random seeds, and we report the average results.


\subsection{RQ1: Overall Performance}
First, as shown in Table~\ref{tab:performance}, comparing ID-based sequential methods, we observe that Mamba4Rec achieves the best performance (e.g., NDCG@10 of 0.0507 on MicroLens), outperforming traditional models like SASRec (0.0428) and GRU4Rec (0.0434). This is because Mamba's selective state space model better captures long-range dependencies in user behavior sequences. In contrast, RNN and Transformer-based models struggle with either gradient issues or quadratic complexity. This demonstrates the importance of efficient sequential modeling in sequential recommendation.

Second, simple multi-modal methods show mixed results compared to their ID-based counterparts. While MMMamba4Rec slightly underperforms Mamba4Rec (0.0492 vs. 0.0507 NDCG@10 on MicroLens), it still maintains relatively strong performance across datasets. This suggests that naive feature concatenation may not always lead to improvements, highlighting the need for more sophisticated multi-modal fusion approaches.

Third, advanced multi-modal methods (NOVA, UniSRec, IISAN, MMMLP) demonstrate clear advantages through their dedicated fusion mechanisms. For instance, MMMLP achieves the best baseline performance with NDCG@5 of 0.0425 on MicroLens, 0.2293 on MovieLens-100K, and 0.0822 on Amazon Beauty. This confirms the value of sophisticated multi-modal modeling strategies.

Finally, our proposed method consistently outperforms all baselines across datasets with significant improvements (1.04-3.34\% relative gains across metrics). Specifically, it achieves NDCG@5 scores of 0.0438, 0.2325, and 0.0843 on MicroLens, MovieLens-100K, and Amazon Beauty respectively. This superior performance can be attributed to three key advantages: (1) the cross-modal expert routing mechanism that dynamically filters and retains high-quality complementary information, (2) the integrated information coordination module utilizing Stein kernel alignment for proper calibration of multi-modal signals, and (3) the theoretically-grounded fusion mechanism that maintains both alignment and uniformity of representations. These innovations enable our method to effectively leverage multi-modal information while maintaining robust sequential modeling capabilities.
\begin{table}[t]
    \caption{Performance comparison under different sequence lengths on Amazon Beauty. \textbf{Bold} numbers denote the best performance, \underline{underlined} numbers represent the second-best results, and $^\ast$ indicates statistical significance at p < 0.05 level using a paired t-test. The bottom row shows the relative improvements of our method over the best baseline.}
    \label{tab:length_compare}
    \resizebox{0.45\textwidth}{!}{
    \begin{tabular}{cccccc}
        \toprule
        Length & Method & NDCG@5 & NDCG@10 & MRR@5 & MRR@10 \\
        \midrule
        \multirow{4}{*}{5-10} 
        & SASRec & 0.0342 & 0.0441 & 0.0293 & 0.0341 \\
        & Mamba4Rec & 0.0369 & 0.0472 & 0.0309 & 0.0347 \\
        & NOVA & 0.0384 & 0.0489 & 0.0318 & 0.0359 \\
        & MMMLP & \underline{0.0407} & \underline{0.0502} & \underline{0.0347} & \underline{0.0386} \\
        & \textbf{Ours} & \textbf{0.0448}$^\ast$ & \textbf{0.0536}$^\ast$ & \textbf{0.0373}$^\ast$ & \textbf{0.0408}$^\ast$ \\
        \midrule
        \multirow{4}{*}{10-30}
        & SASRec & 0.0607 & 0.0756 & 0.0502 & 0.0586 \\
        & Mamba4Rec & 0.0659 & 0.0809 & 0.0538 & 0.0619 \\
        & NOVA & 0.0689 & 0.0842 & 0.0567 & 0.0652 \\
        & MMMLP & \underline{0.0727} & \underline{0.0876} & \underline{0.0602} & \underline{0.0689} \\
        & \textbf{Ours} & \textbf{0.0772}$^\ast$ & \textbf{0.0925}$^\ast$ & \textbf{0.0654}$^\ast$ & \textbf{0.0742}$^\ast$ \\
        \midrule
        \multirow{4}{*}{>30}
        & SASRec & 0.1077 & 0.1254 & 0.0938 & 0.1021 \\
        & Mamba4Rec & 0.1137 & 0.1324 & 0.0983 & 0.1087 \\
        & NOVA & 0.1189 & 0.1418 & 0.1053 & 0.1167 \\
        & MMMLP & \underline{0.1207} & \underline{0.1514} & \underline{0.1218} & \underline{0.1338} \\
        & \textbf{Ours} & \textbf{0.1328}$^\ast$ & \textbf{0.1618}$^\ast$ & \textbf{0.1339}$^\ast$ & \textbf{0.1439}$^\ast$ \\
        \bottomrule
    \end{tabular}
    }
\end{table}
\begin{table}[t]
   \caption{Statistics of interaction sequences in Amazon Beauty.}
   \label{tab:data_stats2}
   \begin{tabular}{ccccc}
       \toprule
       Length Range & Users & Items & Interactions & Sparsity \\
       \midrule
       5-10 & 1,628 & 1,657 & 15,878 & 99.67\% \\
       10-30 & 1,973 & 2,419 & 29,821 & 99.39\% \\
       >30 & 313 & 1,891 & 14,884 & 97.54\% \\
       \midrule
       Total & 3,914 & 2,344 & 60,583 & 99.34\% \\
       \bottomrule
   \end{tabular}
\end{table}
\begin{table}[t]
  \caption{Ablation Study Results}
  \label{tab:ablation}
  \resizebox{0.45\textwidth}{!}{
  \begin{tabular}{ccccc}
      \toprule
      Method & NDCG@5 & NDCG@10 & MRR@5 & MRR@10 \\
      \midrule
      Full Model & 0.0843 & 0.1027 & 0.0722 & 0.0797 \\
      w/o Cross-Attn & 0.0806 & 0.1008 & 0.0681 & 0.0767 \\
      w/o IICM & 0.0795 & 0.0971 & 0.0639 & 0.0705 \\
      w/o MoE & 0.0785 & 0.0956 & 0.0632 & 0.0698 \\
      \bottomrule
  \end{tabular}}
\end{table}

\subsection{RQ2: Analysis of Sequence Length Impact and Modality Contributions}

To analyze how effectively our multimodal information coordination mechanism handles scenarios with different sequence lengths and leverages various modalities, we conduct experiments across different sequence length groups. As shown in Table \ref{tab:length_compare} and \ref{tab:data_stats2}, the results reveal several key findings:

First, in short sequence scenarios (5-10 interactions), our model achieves substantial improvements over baselines (e.g., NDCG@5 of 0.0448 vs. 0.0342 for SASRec, a 30.9\% improvement). This superior performance stems from two aspects: (1) when behavioral patterns are insufficient, our Stein kernel-based alignment mechanism effectively leverages visual and textual signals by learning precise cross-modal correlations, and (2) the IICM module adaptively increases the weights of auxiliary modalities to compensate for limited sequential information, enabling better user preference understanding.
Second, the performance gap remains significant but gradually narrows as sequence length increases (10-30 interactions: 0.0772 vs. 0.0607 NDCG@5; >30 interactions: 0.1327 vs. 0.1077). This trend indicates that while sequential patterns become more reliable with longer histories, our multimodal coordination mechanism automatically adjusts to emphasize behavioral signals while still incorporating complementary visual-textual features when beneficial. The cross-modal expert routing mechanism effectively filters and retains only high-quality signals that complement the sequential patterns.
Third, while the dataset statistics show varying sequence length distributions, the consistent improvements across all length groups suggest that our multimodal coordination mechanism can adaptively balance the utilization of sequential patterns and multimodal features. This adaptive capability ensures robust performance regardless of the available interaction history length, by dynamically adjusting the contribution of each modality based on the sequence characteristics.


\subsection{RQ3: Ablation Study}
To answer \textbf{RQ3}, we conduct comprehensive ablation experiments to evaluate the effectiveness of each key component in FindRec, with results shown in Table~\ref{tab:ablation}. We observe:

(1) Removing the cross-attention mechanism leads to a 4.4\% and 5.7\% drop in NDCG@5 and MRR@5 respectively. This is because without cross-attention, the model loses the ability to dynamically filter and retain high-quality complementary information across modalities, resulting in potential noise interference in multimodal fusion. The performance degradation demonstrates the necessity of selective information fusion.

(2) The removal of the Stein-based information coordination module causes more substantial declines (5.7\% in NDCG@5 and 11.5\% in MRR@5). This occurs because without IICM, the model lacks the mechanism to properly calibrate and align multimodal signals while maintaining their distribution uniformity through entropy maximization, leading to potential feature collapse and suboptimal cross-modal representation learning. The significant impact validates our theoretical insights about the importance of achieving both cross-modal alignment and latent space uniformity in multimodal sequential recommendation.
\begin{figure}[t]
    \centering
    \begin{subfigure}[b]{0.49\linewidth}
        \centering
        \includegraphics[width=\linewidth]{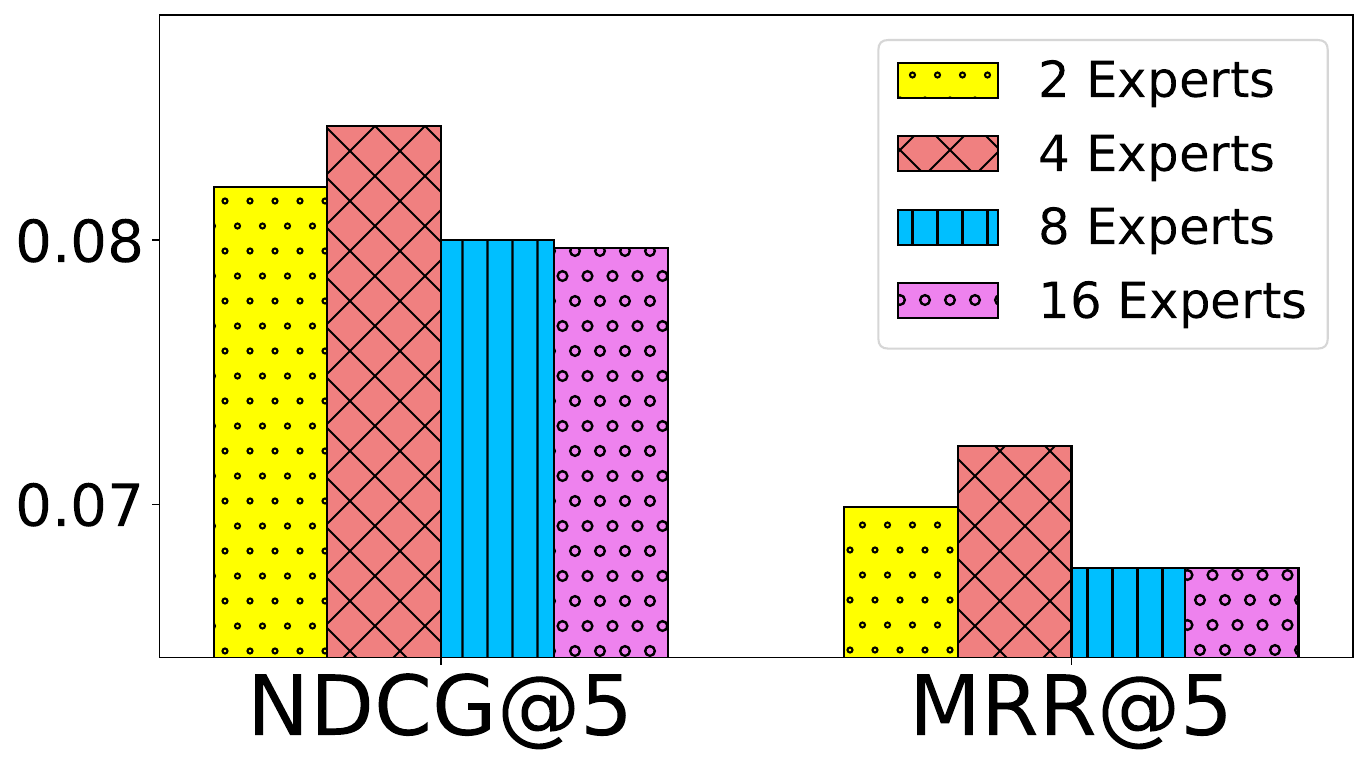}
        \caption{Performance (@5)}
        \label{fig:expert_5}
    \end{subfigure}
    \begin{subfigure}[b]{0.49\linewidth}
        \centering
        \includegraphics[width=\linewidth]{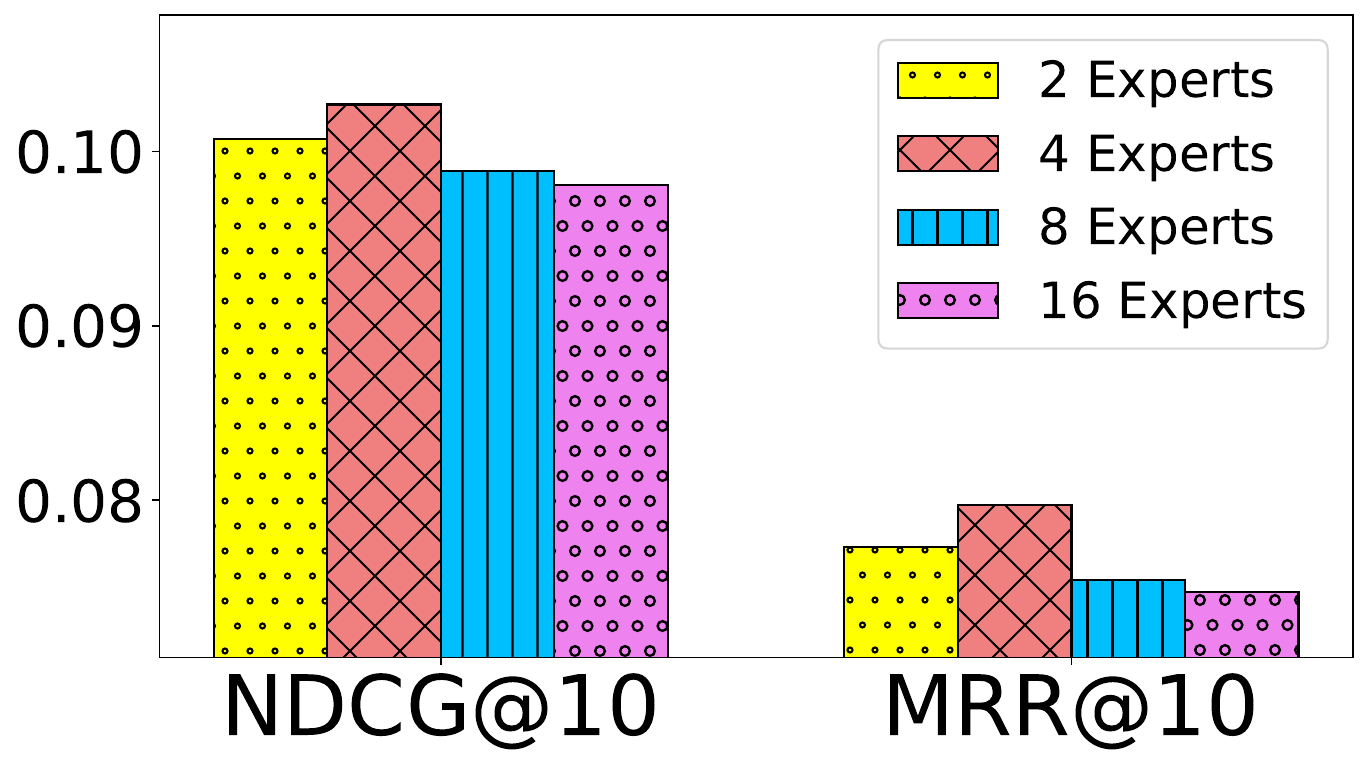}
        \caption{Performance  (@10)} 
        \label{fig:expert_10}
    \end{subfigure}
    \caption{Impact of expert numbers on recommendation}
    \label{fig:expert_analysis}
    \vspace{-5mm} 
\end{figure}
 (3) The Mixture of Experts component shows the most crucial impact, as its removal results in the largest performance drops (6.9\% in NDCG@5 and 12.5\% in MRR@5). This is attributed to the loss of specialized processing for different cross-modal patterns and the adaptive routing mechanism that optimally combines expert knowledge. The substantial performance gap underscores the importance of having multiple specialized experts to handle diverse cross-modal interaction patterns in sequential recommendation.
\subsection{RQ4: Hyperparameter Analysis}
To better understand the impact of different hyperparameters in our model, we conduct ablation studies on several key components. As shown in Figure~\ref{fig:expert_analysis}, we first analyze the impact of expert numbers on model performance. The 4-expert model achieves the best results across all metrics, demonstrating better performance than the 2-expert variant. This improvement stems from 2 experts being insufficient to capture the diverse patterns of cross-modal interactions, while 4 experts provide adequate specialized processing capabilities without introducing excessive computational overhead. However, further increasing the number of experts to 8 and 16 leads to performance degradation, as too many experts result in redundant specialization and increased routing complexity, making the model harder to train effectively and potentially causing expert utilization imbalance. Notably, this trend remains consistent across both @5 and @10 metrics, validating that our choice of 4 experts strikes an optimal balance between model capacity and complexity. These results demonstrate the importance of carefully selecting expert numbers to achieve robust and stable recommendation performance while maintaining computational efficiency.
Due to space limitations, additional hyperparameter analyses on Mamba layer depth, alignment dimension, attention head configuration, and feature alignment spaces are provided in Appendix Sec.~\ref{sec:expmore}.

\section{Related Works}

\subsection{Multimodal Sequential Recommendation}
Early multimodal sequential recommendation, such as MV-RNN~\cite{cui2018mv}, focused on combining latent embeddings with multimodal features via direct concatenation or reconstruction. Subsequent works like TransRec~\cite{wang2024transrec} explored end-to-end pretrained systems, while MLP-based architectures like MMMLP~\cite{liang2023mmmlp} (adapting MLP-Mixer~\cite{tolstikhin2021mlp}) aimed for efficiency and effectiveness by processing multimodal data through dedicated mixing and fusion layers.
More recent efforts have concentrated on sophisticated fusion mechanisms~\cite{bian2023multi} and representation learning. For instance, NOVA~\cite{liu2021noninvasive} utilizes vision-text contrastive learning, and UniSRec~\cite{hou2022towards} offers a unified framework for integrating image, text, and behavior data. Many of these build upon Transformer-based encoders, with variants like sparse Transformers~\cite{li2023strec} addressing efficiency, or employ item-item self-attention as in IISAN~\cite{fu2024iisan} for capturing multimodal similarities. The challenge of effectively integrating diverse information is also echoed in multi-domain recommendation, which uses techniques like hyper-adapters~\cite{li2023hamur} to specialize models. Beyond these, emerging paradigms like prompt-enhanced frameworks~\cite{wang2023plate} offer increased flexibility.

Despite these advancements, many existing methods lack theoretical guarantees for distribution alignment and often rely on static fusion. This makes it difficult to adaptively handle the varying relevance of multimodal features, a limitation FindRec addresses with its Stein kernel-based alignment and dynamic routing, contrasting with dynamic selection in other fields like multi-task learning~\cite{wang2023single}.

\subsection{Stein-Guided Methods.}
The Stein Variational Gradient Descent (SVGD) technique~\cite{liu2016stein} offers a robust framework for distribution alignment and Bayesian inference, optimizing particles via kernel-based deterministic transformations. Advances like projected SVGD~\cite{chen2020projected} have extended its applicability to complex distributions. Stein methods have proven effective in various alignment tasks: MVEB~\cite{wen2024mveb} uses them for multi-view learning by aligning views with von Mises-Fisher kernels and maximizing embedding entropy, while DisAlign~\cite{liu2021leveraging} applies Stein path alignment for cross-domain recommendation. To our knowledge, FindRec is the first to introduce Stein's methods to multimodal sequential recommendation, proposing novel IICM mechanisms that leverage Stein kernels for calibrating multimodal signals and enabling effective information coordination.

\section{Conclusion}
This paper introduces FindRec, a novel framework that addresses fundamental challenges in multimodal sequential recommendation through a principled ``flow-control-align-output'' paradigm. Our framework makes three key contributions to the field. We achieve theoretically guaranteed cross-modal consistency through the Stein kernel-based Integrated Information Coordination Module (IICM) while preserving modality-specific characteristics. This innovation resolves the long-standing challenge of balancing alignment precision with information preservation. Our dynamic cross-modal expert routing mechanism also effectively manages varying degrees of feature relevance, demonstrating superior performance in handling noisy inputs and long-tail interactions. 
The principles underlying FindRec also open avenues for extensions to even more complex recommendation ecosystems, such as whole-chain recommendation scenarios~\cite{zhao2020whole}.
A detailed discussion of limitations and other promising future directions, including advanced evaluation methodologies~\cite{zhao2023kuaisim}, can be found in Appendix Sec.~\ref{sec:Limitations}.

\begin{acks}
This research was partially supported by Research Impact Fund (No.R1015-23), Collaborative Research Fund (No.C1043-24GF), Huawei (Huawei Innovation Research Program, Huawei Fellowship), Tencent (CCF-Tencent Open Fund, Tencent Rhino-Bird Focused Research Program), Alibaba (CCF-Alimama Tech Kangaroo Fund No. 2024002), Ant Group (CCF-Ant Research Fund), and Kuaishou. We also thank Dr. Wen Liangjian for his valuable discussion. 
\end{acks}

\newpage

\bibliographystyle{ACM-Reference-Format}
\balance
\bibliography{main}

\begin{appendices}

\section{More Hyperparameter Experimets}
\label{sec:expmore}

Our extensive hyperparameter analysis reveals several important findings. Table~\ref{tab:lamb} shows the impact of the balance parameter $\lambda$, where the model achieves optimal performance at $\lambda=1e-3$ (NDCG@5: 0.0843, MRR@5: 0.0722) compared to other values. When $\lambda$ is too small (1e-4), the model insufficiently emphasizes cross-modal alignment, leading to inadequate feature fusion. Conversely, larger values (1e-1) over-emphasize alignment at the expense of preserving modality-specific characteristics. This validates our theoretical framework's emphasis on maintaining a delicate balance between cross-modal alignment and modality preservation.
 
The analysis of Mamba layer depth in Table~\ref{tab:lamb} demonstrates that the 2-layer architecture consistently outperforms other configurations (NDCG@5: 0.0843, MRR@5: 0.0722). While single-layer models (NDCG@5: 0.0825) lack sufficient capacity to capture complex sequential dependencies, deeper architectures (3 and 4 layers) introduce optimization challenges and increased overfitting risks, particularly given the sparse nature of recommendation data. This finding underscores the importance of architectural efficiency in sequential modeling, suggesting that moderate-depth architectures can effectively capture necessary sequential patterns while maintaining computational traceability.

The investigation of alignment dimension (Table~\ref{tab:layer1}) reveals a strong correlation between dimensional capacity and model effectiveness. The performance metrics demonstrate consistent enhancement with increased dimensions, culminating in optimal results at 512 dimensions (NDCG@10: 0.1044, MRR@10: 0.0807). This observation substantiates our hypothesis that higher-dimensional alignment spaces facilitate more nuanced multimodal feature interactions and preserve modal-specific information structures. The empirical evidence strongly supports our architectural decisions in balancing model expressiveness and computational efficiency.

Furthermore, we conduct comprehensive analyses on attention mechanism configurations and feature alignment spaces. As illustrated in Table~\ref{tab:head_analysis}, the attention head configuration exhibits a non-monotonic relationship with model performance. The 8-head architecture achieves superior performance (NDCG@5: 0.0843, MRR@5: 0.0722), while further head expansion to 16 results in performance deterioration, particularly manifested in the significant MRR@5 degradation to 0.0501. This phenomenon aligns with our theoretical analysis that excessive attention heads may lead to redundant feature interactions and optimization difficulties.

\begin{table}[h!]
\centering
\caption{Performance comparison with different $\lambda$ values}
\label{tab:lamb}
\begin{tabular}{cccccc}
\hline
$\lambda$ & NDCG@5 &  NDCG@10 & MRR@5 & MRR@10 \\
\hline
1e-4 & 0.0826 & 0.1015 & 0.0713 & 0.0790 \\
1e-3 & 0.0843 & 0.1027 & 0.0722 & 0.0797 \\
1e-2 & 0.0797 & 0.0989 & 0.0663 & 0.0734 \\
1e-1 & 0.0816 & 0.0993 & 0.0682 & 0.0752 \\
\hline
\end{tabular}
\end{table}

\begin{table}[h!]
   \centering
   \caption{Analysis of Mamba Layer Numbers}
   \begin{tabular}{c|cccc}
       \hline
       Layers & NDCG@5 & NDCG@10 & MRR@5 & MRR@10 \\
       \hline
       1 & 0.0825 & 0.1005 & 0.0709 & 0.0791 \\
       2 & 0.0843 & 0.1027 & 0.0722 & 0.0797 \\
       3 & 0.0803 & 0.0986 & 0.0678 & 0.0761 \\
       4 & 0.0794 & 0.0962 & 0.0665 & 0.0742 \\
       \hline
   \end{tabular}
\end{table}
\begin{table}[h!]
   \centering
   \caption{Analysis of Alignment Dimemsion}
   \label{tab:layer1}
   \begin{tabular}{c|cccc}
       \hline
       Dimensions & NDCG@5 & NDCG@10 & MRR@5 & MRR@10 \\
       \hline
       128 & 0.0785 & 0.0981 & 0.0669 & 0.075 \\
       256 & 0.0804 & 0.0989 & 0.0687 & 0.0761 \\
       512 & 0.0838 & 0.1044 & 0.0723 & 0.0807 \\
       \hline
   \end{tabular}
\end{table}

\begin{table}[h!]
   \centering
   \caption{Analysis of Attention Heads}
   \label{tab:head_analysis}
   \begin{tabular}{c|cccc}
       \hline
       Heads & NDCG@5 & NDCG@10 & MRR@5 & MRR@10 \\
       \hline
       2 & 0.0821 & 0.1001 & 0.0682 & 0.0764 \\
       4 & 0.0808 & 0.1016 & 0.0694 & 0.0778 \\
       8 & 0.0843 & 0.1027 & 0.0722 & 0.0797 \\
       16 & 0.0823 & 0.0957 & 0.0687 & 0.0776 \\
       \hline
   \end{tabular}
\end{table}
These results collectively validate our architectural design choices and provide practical guidelines for implementing similar architectures in real-world recommendation systems, emphasizing the importance of careful hyperparameter tuning in balancing model capacity and learning effectiveness.

\section{Limitations and Future Work}
\label{sec:Limitations}
While FindRec demonstrates promising results, we acknowledge several limitations that suggest important directions for future research. The primary limitation lies in our current modeling of temporal patterns, which does not explicitly consider periodicity and cyclical patterns in user behaviors. Although our Mamba-based sequential modeling captures general temporal dependencies, it may miss important seasonal trends, daily/weekly cycles, and periodic preference shifts that are common in real-world recommendation scenarios. For instance, in e-commerce settings, user preferences often exhibit strong weekly patterns (weekend vs. weekday shopping behaviors) and seasonal variations (holiday shopping trends), which our current model might not fully capture. Additionally, our cross-modal feature extraction treats all temporal contexts equally, potentially overlooking how the importance of different modalities (visual, textual) varies across different time periods or seasonal contexts. For example, visual features might be more crucial during fashion shopping seasons, while textual descriptions could be more important during technical product launches.

These limitations motivate several promising directions for future research. 
First, incorporating explicit periodic components into our sequential modeling mechanism could better capture cyclical patterns in user behaviors. This could involve integrating Fourier transformations or dedicated periodic attention mechanisms to model both short-term and long-term periodic dependencies. The enhanced temporal modeling would allow the system to better predict user preferences based on historical periodic patterns, such as weekend shopping habits or seasonal buying trends. 

Second, developing time-aware multimodal feature extraction that considers temporal context when processing visual and textual information could enhance the model's ability to capture evolving cross-modal patterns. This would involve designing adaptive feature extraction mechanisms that dynamically adjust the processing of different modalities based on temporal context, potentially improving the model's ability to capture time-varying cross-modal relationships.

Third, exploring dynamic routing mechanisms that adapt to different temporal contexts could help the model better handle varying importance of different modalities across time periods. For instance, the expert routing mechanism could be enhanced to consider temporal factors when determining the importance of different modalities, allowing for more nuanced handling of multimodal information across different temporal contexts. This could involve developing time-sensitive gating mechanisms or temporal attention layers that modulate the contribution of different experts based on temporal patterns. These enhancements would allow FindRec to better model the complex temporal dynamics inherent in real-world recommendation scenarios while maintaining its strengths in multimodal information coordination and interpretability. 

Fourth, future work could explore the integration of external temporal information, such as holiday calendars or event schedules, to further enhance the model's ability to capture meaningful temporal patterns in user behaviors and item preferences.

Beyond these direct extensions related to temporal dynamics and multimodal fusion, several broader avenues warrant investigation.
One such direction involves applying and potentially adapting FindRec's principles to more complex and holistic recommendation ecosystems. For example, its utility in whole-chain recommendation systems~\cite{zhao2020jointly}, which consider the entire user journey from discovery to conversion and beyond, could be explored. Similarly, investigating scenarios that require joint optimization of recommendation with other related objectives, such as advertising effectiveness~\cite{zhao2020jointly}, could lead to models with greater overall system utility.

Furthermore, advancing the evaluation methodologies for multimodal sequential recommenders remains a critical area. While experiments on static real-world datasets are invaluable, supplementing these with evaluations in comprehensive and controllable simulation environments~\cite{zhao2021usersim, zhao2023kuaisim} can provide deeper, more reproducible insights into algorithmic behavior. Such simulators can facilitate the study of long-term user engagement, the impact of recommendation strategies over extended periods, and the robustness of models to various data shifts and user interaction patterns. This would contribute to a more thorough understanding and more reliable deployment of sophisticated recommendation models.
\end{appendices}

\end{document}